\documentclass[aps, prd, preprint, amsfonts,
  amssymb, amsmath, showpacs, letterpaper,nofootinbib]{revtex4-1}

 \usepackage{bm,slashed}

\usepackage{color,graphicx}

\begin{document}

\title{Symmetron Scalar Fields: \\
Modified Gravity, Dark Matter or Both?}

\author{Clare Burrage}
\email{clare.burrage@nottingham.ac.uk}
\affiliation{School of Physics and Astronomy, University of Nottingham,\\ Nottingham NG7 2RD, United Kingdom}

\author{Edmund J. Copeland}
\email{edmund.copeland@nottingham.ac.uk}
\affiliation{School of Physics and Astronomy, University of Nottingham,\\ Nottingham NG7 2RD, United Kingdom}

\author{Christian K\"{a}ding}
\email{christian.kading@nottingham.ac.uk}
\affiliation{School of Physics and Astronomy, University of Nottingham,\\ Nottingham NG7 2RD, United Kingdom}

\author{Peter Millington}
\email{p.millington@nottingham.ac.uk}
\affiliation{School of Physics and Astronomy, University of Nottingham,\\ Nottingham NG7 2RD, United Kingdom}


\begin{abstract}
Scalar fields coupled to gravity through the Ricci scalar have been considered  both as dark matter candidates and as a possible modified gravity explanation for galactic dynamics. It has recently been demonstrated that the dynamics of baryonic matter in disk galaxies may be explained, in the absence of particle dark matter, by a symmetron scalar field that mediates a fifth force. The symmetron provides a concrete and archetypal  field theory within which to  explore how large a role modifications of gravity can play on galactic scales.  In this article, we extend these previous works by asking whether the same symmetron field  can explain the difference between the baryonic and lens masses of galaxies through a modification of gravity.   We consider the possibilities for minimal modifications of the model and find that this difference cannot be explained entirely by the symmetron fifth force without extending the field content of the model. Instead, we are pushed towards a regime of parameter space  where one scalar field both mediates a fifth force and stores enough energy density  that it also contributes to the galaxy's gravitational potential  as a dark matter component, a regime which remains to be fully explored. 

\end{abstract}

\maketitle

\section{Introduction}
A suite of cosmological observations, from the Cosmic Microwave Background (CMB) to galactic rotation curves \cite{Bertone:2016nfn,Bergstrom:2000pn}, provides compelling evidence that there must be new physics in the universe beyond the Standard Model of particle physics.  It is remarkable that the introduction of an additional pressureless perfect fluid, combined with a cosmological constant, is sufficient to allow the standard $\Lambda$CDM model of cosmology to agree with all observations to date. However, the composition of the dark matter perfect fluid is currently unclear.   It is common to assume that dark matter has a particulate nature, but the mass of the  dark matter candidate is effectively undetermined, with possibilities ranging from ultra-light axions, $m \sim 10^{-22} \mbox{ eV}$ \cite{Marsh:2015xka}, to intermediate-mass black holes, $m \sim 10^2 M_{\odot}$ \cite{Carr:2016drx}.  A direct detection of a dark matter particle still eludes us, and, until its particle nature has been determined, the question of whether modified gravity could also form part of the new physics required to explain observations commonly attributed to particle dark matter remains open. 

In this work, we will consider a scalar-tensor modification of gravity known as the symmetron model. The symmetron is a canonical scalar field with a spontaneous symmetry breaking potential, which couples universally to the trace of the matter energy-momentum tensor \cite{Hinterbichler:2010es,Hinterbichler:2011ca} (see also Refs.~\cite{Dehnen:1992rr,Gessner:1992flm,Damour:1994zq,Pietroni:2005pv,Olive:2007aj} for related earlier work).  The form of the symmetron potential and coupling mean that, in regions of high density, the $\mathbb{Z}_2$ symmetry of the theory is restored and, in regions of low density, it is spontaneously broken.  As the dominant effective coupling constant between the scalar and matter is proportional to the vacuum expectation value (vev) of the scalar field, we find that, when the local density is high enough to restore the symmetry, the scalar fifth force decouples from matter.
This environmental dependence  allows the theory to evade the constraints from fifth-force searches and local tests of gravity, and this effect is known as screening. (For a review of screening in scalar-tensor theories of gravity, see Ref.~\cite{Joyce:2014kja}.)

The symmetron \cite{Hinterbichler:2010es,Hinterbichler:2011ca} was first introduced as a possible explanation for dark energy and the late-time accelerated expansion of the universe \cite{Copeland:2006wr,Clifton:2011jh}, although it still requires the presence of a cosmological constant term to match observations. It also suffers from the same fine-tuning problems  as the cosmological constant to explain the measured value of this term. It has, however,  recently been shown\footnote{By some of the authors of this article.} that a symmetron scalar field, and its associated fifth force, could explain both the rotation curves and stability of disk galaxies \cite{Burrage:2016yjm} and the motion of stars perpendicular to the plane of the Milky Way disk \cite{OHare:2018ayv}. As such,  the symmetron has the potential to be an interesting modified gravity alternative to particle dark matter, although it should be made clear that particle dark matter passes many observational tests for which the effects of the symmetron have not yet been computed.  Modified gravity alternatives to particle dark matter, such as MOND \cite{Milgrom:2016uye}, MOG \cite{Moffat:2016ikl} and TeVeS \cite{Skordis:2009bf}, have a long history. In contrast to these models, the symmetron model adds only a single scalar field to the standard models of particle physics and cosmology, within a well-defined effective field theory framework.  In fact, (restricting to dimension-four operators) the symmetron is nothing other than a Higgs-portal model to a light scalar sector with spontaneous symmetry breaking \cite{Burrage:2018dvt}.  Viewed from this  angle, the symmetron model is  an extension of the Standard Model of particle physics, coupled minimally to Einstein gravity, but involving an additional light scalar field that mixes with the would-be standard model Higgs field.  

Scalar fields  have a long  history of being studied as  particle dark matter candidates.  Heavy scalars can play the role of dark matter, produced thermally in the early universe,  if they are neutral under the standard model gauge group \cite{Silveira:1985rk,McDonald:1993ex,Burgess:2000yq}.  Very light scalars can also be dark matter when they are sufficiently weakly coupled that they are not thermally produced in the early universe \cite{Turner:1983he,Press:1989id,Sin:1992bg,Hu:2000ke,Goodman:2000tg,Peebles:2000yy,Arbey:2001qi,Lesgourgues:2002hk,Amendola:2005ad,Schive:2014dra,Hui:2016ltb}. The  Compton wavelength of these scalars is order the size of a typical galaxy, $\sim 10 \mbox{ kpc}$, and so they form a coherent condensate around the galaxy. In other cases, they can give rise to soliton-like configurations in the galactic core~\cite{Chavanis:2011zi,Chavanis:2011zm,Marsh:2015wka,Chen:2016unw,Bar:2018acw}. Recent work has begun to study explicit couplings between scalar dark matter and the Ricci scalar, and  this  non-minimal coupling provides a natural explanation for the production of such dark matter in the early universe \cite{Markkanen:2015xuw, Ema:2016hlw,Ema:2018ucl,Fairbairn:2018bsw,Alonso-Alvarez:2018tus}. 

In this work, and so as to be clear about the distinct types of phenomenology arising from the symmetron model,   we will say that the scalar behaves as dark matter, if the background scalar configuration contributes significantly to the total energy density of a galaxy.  If this is not the case, but perturbations in the scalar field mediate an additional force, we will say that the theory is in the modified gravity regime. 
The main difference between the symmetron model and other  scalar theories typically considered as dark matter is the important role played by non-linear behaviour and long-range coherent effects that allow the symmetron  to mediate a fifth force on galactic scales while still being screened in the Solar System.

The symmetron model bridges both dark matter and modified gravity explanations for galactic dynamics, and can move between the two depending on the choice of parameters in the potential and the strength of the coupling to matter.\footnote{We note that, elsewhere in its  parameter space, the symmetron can play a role during inflation \cite{Dong:2013swa,Brax:2014baa,Antusch:2014qqa}.} In this way, the symmetron provides an important field-theoretic archetype. Previous work has shown that the symmetron fifth force can explain the internal dynamics of galaxies, without contributing significantly to the total mass of the galaxy.  This begs the question of whether the symmetron model, acting as a modification of gravity, can be consistent with observations of gravitational lensing around galaxies, where the `lens mass' of the galaxy is also found to be larger than can be accounted for with just the galactic baryonic mass. First constraints on the strength of the coupling between an additional modified-gravity scalar, which is responsible for galactic dynamics in the absence of particle dark matter, and photons were derived in Ref.~\cite{Dai:2018fxc} from observations of the lens system SDSS J2141-0001. It was found that  the effect of the modified-gravity scalar on the lensing of photons must be of the same order as its effect on the motion of matter within the galaxy.

In this work, we ask precisely this question: Considering the minimal possibilities in detail, can the symmetron model explain both the internal dynamics of a galaxy and how it lenses distant sources? In Section \ref{sec:symm}, we introduce the symmetron model and, in Section \ref{sec:intra}, we review the previous work, arguing that a symmetron-mediated fifth force can explain the internal dynamics of galaxies without the need for a dark matter component. In Section \ref{sec:lens}, we consider all of the possible ways in which photon geodesics could be modified by a symmetron field acting as a modification of gravity.  Firstly, we check that, in the regime of interest, the symmetron does not contribute significantly to the overall energy budget of the galaxy. We then consider the consequences in turn  of introducing axion-like couplings to the electromagnetic field, a photon mass, and a disformal coupling. We find that none of these minimal extensions of the symmetron can explain gravitational lensing by galaxies in the absence of dark matter.  We conclude in Section \ref{sec:conc} and consider some possibilities for further study of the  symmetron models as a hybrid dark matter - modified gravity model.

Throughout this work, we choose the mostly plus sign convention for the metric and set $c=\hbar=1$. 

\section {The symmetron model}
\label{sec:symm}
As a scalar-tensor theory, the original symmetron model is only defined up to conformal transformations and  associated field redefinitions. The two most common formulations are the Jordan frame, where the scalar couples explicitly to the Ricci scalar and not directly to matter fields, and the Einstein frame where there is no non-minimal coupling to gravity but  a direct coupling to matter fields \cite{Will:2014kxa}. 
The symmetron model consists of a non-minimally coupled scalar field whose evolution, in the Einstein frame, is governed by an effective potential
\begin{equation}
V_{\rm eff}(\varphi)\ =\ \frac{1}{2}\bigg(\frac{\rho}{M^2}\:-\:\mu^2\bigg)\varphi^2\:+\:\frac{1}{4}\,\lambda\,\varphi^4\;,
\label{eq:sympot}
\end{equation}
where $\mu^2>0$ corresponds to  a constant (tachyonic) mass, $\lambda>0$ is a dimensionless constant, and $\rho$ is the non-relativistic energy density of matter. 

The density dependent term in the effective potential, Eq. (\ref{eq:sympot}), arises because matter fields move on geodesics of 
 the Jordan-frame metric $g_{\mu\nu}=A^2(\varphi)\tilde{g}_{\mu\nu}$, where $\tilde{g}_{\mu\nu}$ is the Einstein frame metric. The coupling function $A(\varphi)$ has the form
\begin{equation}
A(\varphi)\ =\ 1\:+\:\frac{\varphi^2}{2M^2}\:+\:\mathcal{O}\bigg(\frac{\varphi^4}{M^4}\bigg)\;,
\label{eq:scalarcoupling}
\end{equation}
and the constant energy scale $M$ defines the strength of  the
 interactions between the scalar and matter fields. 
This model should be considered an effective field theory, valid up to the scale $M$. Whilst the purely scalar sector consists only of operators up to mass dimension four, the coupling to matter introduces higher-dimension operators, which are non-renormalisible, reflecting the non-renormalisability of the modified gravitational sector in the Jordan frame.

As stated above, matter particles move on geodesics of the Jordan frame metric $g_{\mu\nu}$. In terms of the Einstein frame metric, this can be understood as matter particles feeling an additional fifth force mediated by the symmetron scalar. A unit-mass test particle is subject to a fifth force of the form (see, e.g., Ref.~\cite{Brax:2012nk})
\begin{equation}
\label{eq:symforce}
\vec{F}_{\rm sym}\: =\: -\,\vec{\nabla}\,\ln\,A(\varphi)\: \approx\: -\,\frac{\varphi}{M}\,\vec{\nabla}\,\frac{\varphi}{M}\;.
\end{equation}
In regions of sufficiently high density, where $\rho/M^2>\mu^2$, the minimum of the effective potential in Eq. (\ref{eq:sympot}) lies at the origin, $\varphi=0$, and the $\mathbb{Z}_2$  symmetry of the theory is restored. In such regions, the coupling strength $\varphi/M$ therefore goes to zero, and the fifth force is {\it screened}.
In regions of sufficiently low density, where $\rho/M^2  \ll\mu^2$, the $\mathbb{Z}_2$ symmetry is broken, and the symmetron field acquires a non-zero vev $\varphi\approx \pm\,v=\pm\,\mu/\sqrt{\lambda}$. In such regions, spatial gradients in the symmetron field will give rise to a fifth force, with coupling strength $v/M$.  

A variant of the symmetron model can also be realised by the Coleman-Weinberg mechanism~\cite{Coleman:1973jx} of spontaneous symmetry breaking (see Ref.~\cite{Burrage:2016xzz}). While the precise shape of the symmetry-breaking potential differs from that in Eq.~\eqref{eq:sympot}, the general arguments presented in this work carry over to this alternative construction.

As the symmetron model is designed to avoid local tests of gravity, constraints on the model parameters are weak.  They arise from Parametrized Post-Newtonian (PPN) bounds in the Solar System \cite{Hinterbichler:2010es,Hinterbichler:2011ca,OHare:2018ayv}, from precision tests of gravity  in the laboratory \cite{Upadhye:2012rc,Burrage:2016rkv,Brax:2016wjk,Brax:2017hna,Brax:2017xho,Brax:2018zfb,Llinares:2018mzl}, and from astrophysical probes \cite{Davis:2011pj,Sakstein:2015oqa,Sakstein:2017pqi}. These constraints are reviewed in full in Ref.~\cite{Burrage:2017qrf}.   The remaining parameter space, however, is vast.  The expected values for the model parameters vary depending on the motivation for studying the symmetron, in the initial formulation, $\mu$ and $M$ were chosen so that the critical density for the symmetry breaking transition $\rho_{\rm crit} = \mu^2 M^2$ was of order the critical density of the universe today, i.e.\ $\rho_{\rm crit} \sim H_0^2 M_{\rm Pl}^2$. This would mean that  a phase transition in the late universe could explain the onset of dark energy domination.  If the scalar field is thought to arise from a modification of gravity then one might expect that $M\sim M_{\rm Pl}$, where $M_{\rm Pl}$ is the reduced Planck mass.  However,  we are not required to make either of these assumptions about the parameters of the model. In this work, we do not place any theory priors on the expected values of the parameters.

\section{Intra-galactic effects of conformally coupled symmetrons}
\label{sec:intra}
For an isolated galaxy, the symmetron field profile will interpolate between $\varphi= v$ in the cosmological vacuum and a smaller value at the center of the galaxy $\varphi(0) <v$.  As the central density of the galaxy is increased, this value will asymptote to zero.  It is clear from Eq.~(\ref{eq:symforce}) that the strength of the fifth force experienced by individual particles or stars within the galaxy depends both on the value and the gradient of $\varphi$ at their respective position.

In Ref.~\cite{Burrage:2016yjm}, it was shown that this fifth force may be sufficient to explain the rotation curves of disk galaxies without the need for particle dark matter. To see how this is possible,  we consider the SPARC data set~\cite{Lelli:2016zqa}, the analysis of which showed that, in disk galaxies, the observed centripetal accelerations ($g_{\rm obs}$) and those predicted from the baryonic component alone ($g_{\rm bar}$) follow the empirical relation \cite{McGaugh:2016leg}
\begin{equation}
\label{eq:relrel}
g_{\rm obs}\ =\ \frac{g_{\rm bar}}{1-e^{-\sqrt{g_{\rm bar}/g_{\dag}}}}\ =\ g_{\rm bar}\:+\:\frac{g_{\rm bar}}{e^{\sqrt{g_{\rm bar}/g_{\dag}}}-1}\:\;,
\end{equation}
where $g_{\dag}=1.20\pm 0.02 \mathrm{(rand.)}\pm 0.24\mathrm{(sys.)}\!\times\! 10^{-10}\, {\rm ms^{-2}}$.
To see if the symmetron may be able to explain this effect, we approximate the galaxies as thin disks, of uniform  density  and height $h$.  Then the symmetron force in Eq.~\eqref{eq:symforce} contributes a centripetal acceleration
\begin{equation}
\label{eq:match1}
g_{\rm sym}(r)\ =\ \frac{1}{2}\,\frac{\rm d}{{\rm d} r}\,\bigg(\frac{\varphi(r)}{M}\bigg)^{\! 2}\;,
\end{equation}
To produce the  correlation in Eq.~\eqref{eq:relrel}, the symmetron field must give rise to an acceleration of the form
\begin{equation}
\label{eq:match2}
g_{\rm sym}(r)\ =\ \frac{g_{\rm bar}(r)}{e^{\sqrt{g_{\rm bar}(r)/g_{\dag}}}-1}\;.
\end{equation}

\begin{figure}[!t]
\centering
        \includegraphics[width=0.6\textwidth]{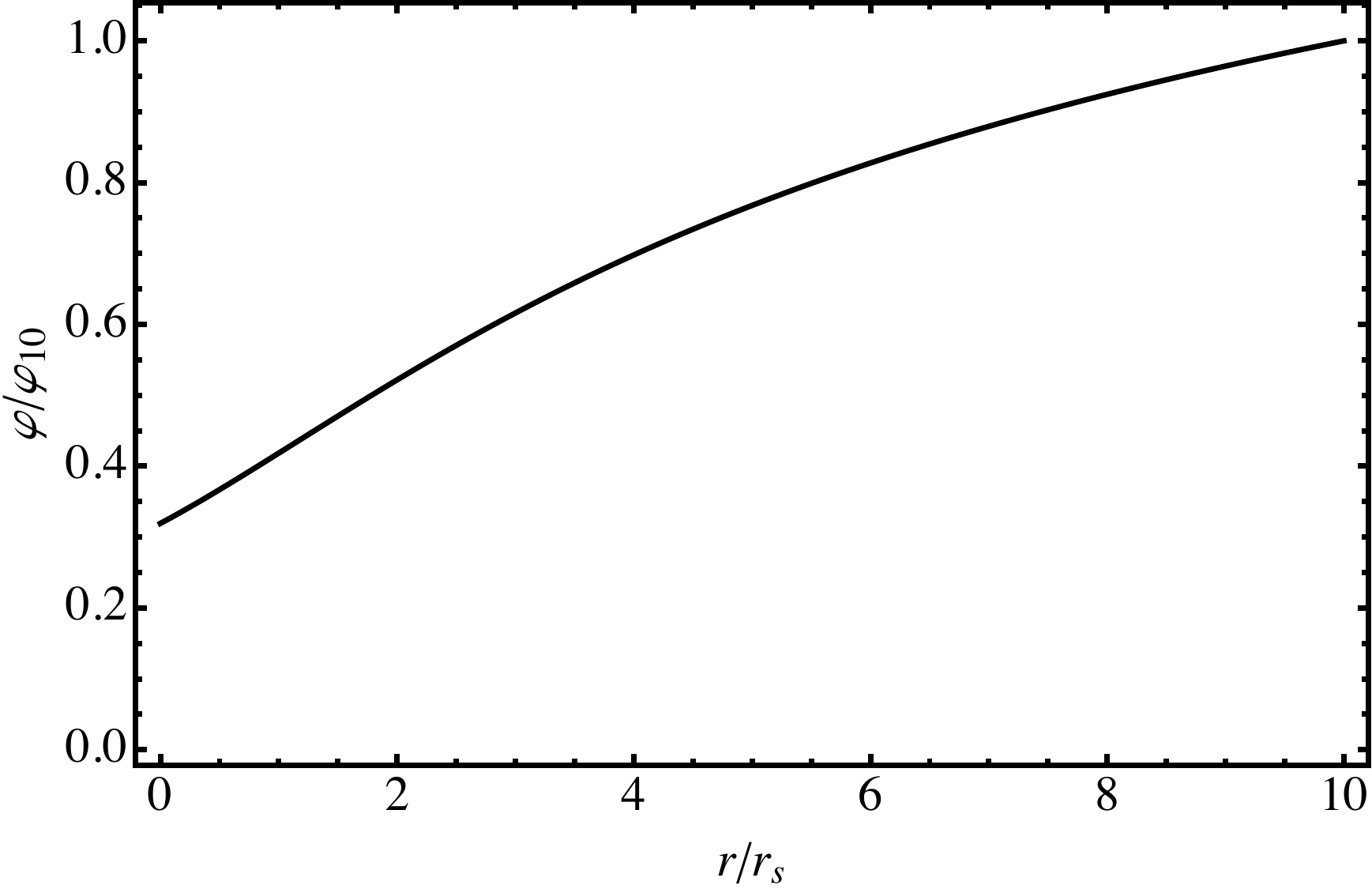}
  \caption{The symmetron profile $\varphi$ of Eq.~\eqref{eq:phiint}.  The associated fifth force would reproduce the empirical relationship of Eq.~(\ref{eq:relrel}) to describe the rotation curves for a simplified model of a disk galaxy. We have normalized to the value of the field  at $r/r_s=10$ ($\varphi_{10}$) with boundary condition $\varphi_0/M=10^{-3}$. Figure reproduced with permission from Ref.~\cite{Burrage:2016yjm}.}
	\label{fig:pete}
\end{figure}

The symmetron profile is sourced by the distribution of baryons in the galaxy. We assume an exponential disk profile for the surface mass density of the galaxy $\Sigma(r)\ =\ \Sigma_0\,e^{-r/r_s}$,
so that the total baryonic mass within a radius $r$ is given by
\begin{equation}
\mathcal{M}_{\rm bar}(r)\  =\ \mathcal{M}_0\left[1-e^{-r/r_s}\left(1+\frac{r}{r_s}\right)\right]\;.
\end{equation}
Here, the total baryonic mass of the galaxy is $\mathcal{M}_0=2\pi r_s^2\Sigma_0$  and $r_s$ is its scale length. For convenience, we define $x\equiv r/r_s$ and
\begin{equation}
f(x)\ \equiv\ \frac{f_0}{x} \big[1\:-\:e^{-x}\big(1+x\big)\big]^{\tfrac{1}{2}}\;,\quad f_0\ =\ \Big(\frac{G\mathcal{M}_0}{g_{\dag} r_s^2}\Big)^{\tfrac{1}{2}}\;.
\end{equation}
Then, using the coarse approximation that $g_{\rm bar} \approx  (G \mathcal{M}_{{\rm bar}}(r))/r^2$,
we see that we require
\begin{equation}
\label{eq:phiint}
\bigg(\frac{\varphi}{M}\bigg)^{\! 2}\ \approx \ \bigg(\frac{\varphi_0}{M}\bigg)^{\!2}\:+\:2\,\frac{g_{\dag}r_s}{c^2}\!\int_{0}^x\!{\rm d}x'\;\frac{f^2(x')}{e^{f(x')}-1}\;.
\end{equation}
A galaxy with a mass and scale length comparable to the Milky Way ($\mathcal{M}_0\approx6\times10^{11}\ {\rm M}_{\odot}$ and $r_s\approx 5\ {\rm kpc}$) will have  $f_0\approx 5$. The integral in Eq.~\eqref{eq:phiint} appears to diverge  as $x \to \infty$, however we recall that  the identification in Eq.~\eqref{eq:match2} need only hold out to a finite radius, which would provide a cut-off for the integral in Eq.~\eqref{eq:phiint}. The form of this field profile is plotted in Fig.~\ref{fig:pete}.

A concrete, but simplified,  example was simulated in Ref.~\cite{Burrage:2016yjm} (related earlier work can be found in Refs.~\cite{Dehnen:1992rr,Gessner:1992flm}) with $M= M_{\rm Pl}/10$, $v/M=1/150$ and $\mu = 3 \times 10^{-39} \mbox{ GeV}$, to show that symmetron profiles of the required form, naturally arise within the set of galaxies in the SPARC data set.  Additionally, this symmetron component was shown to explain both the stability of the disk galaxy to the formation of bars, and the radial acceleration relation of \cite{Janz:2016nwx,McGaugh:2016leg}. This choice of parameters is in conflict with the bounds coming from constraints on the PPN parameters. However, it was noticed in this work that there is a degeneracy in the $\mu,M$ parameter space where the effective mass of the symmetron in the galaxy approximately vanishes $m_{\rm eff}^2 =  \rho_{\rm gal}/M^2 -\mu^2 \approx 0$.  So it is expected that there is a choice of parameters that can simultaneously explain the galactic dynamics and be compatible with local tests of gravity, although numerical simulations in this non-linear regime are challenging.

\begin{figure}[!pth]
\centering
        \includegraphics[width=0.7\textwidth]{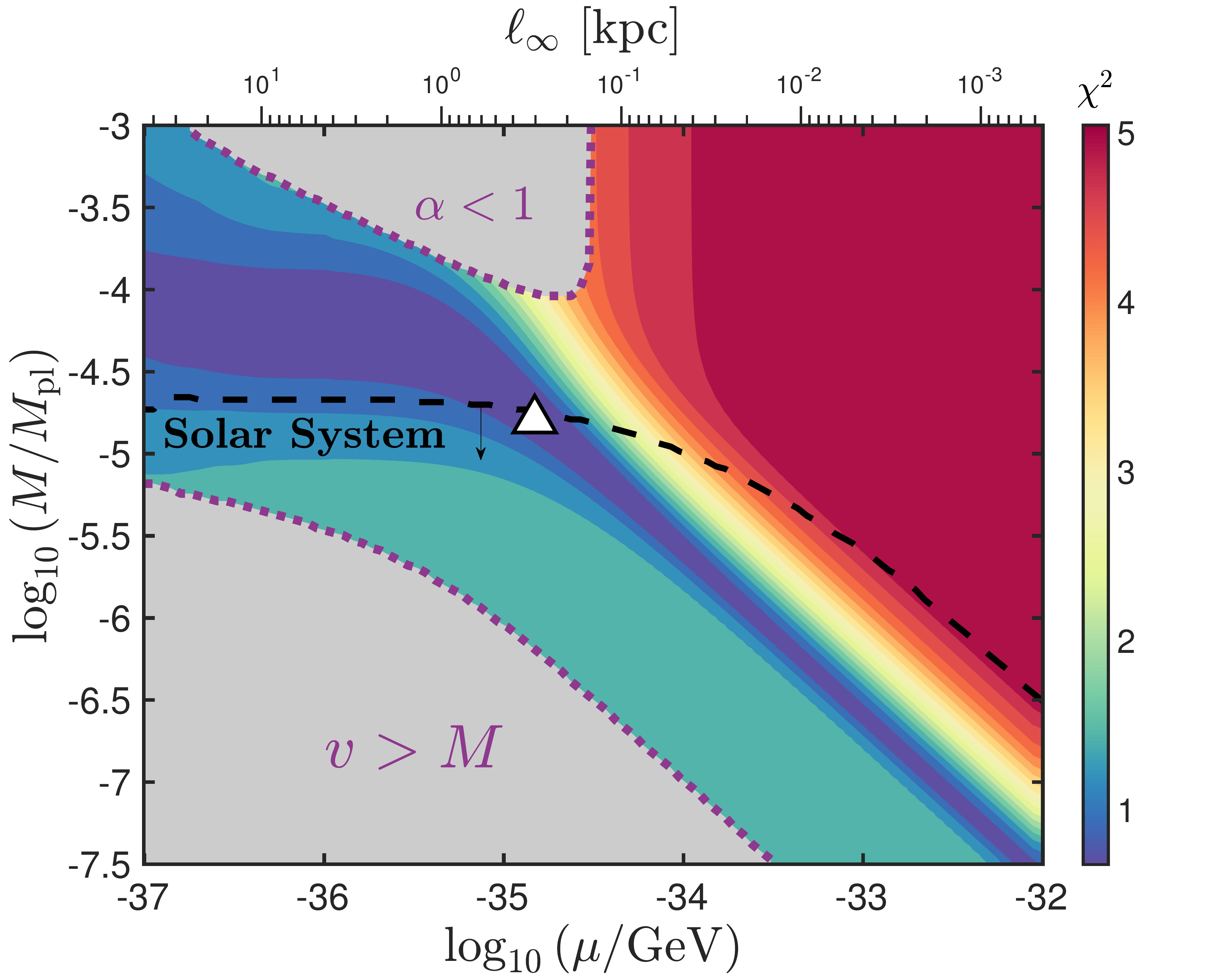}
  \caption{Reduced $\chi^2$ value for the fit of the symmetron modified gravity model to  the velocity dispersion of stars perpendicular to the plane of the galaxy for the mock data set of Ref.~\cite{Read:2014qva}.   Values are plotted as  a function of the symmetron parameters $M$, and $\mu$ and the Compton wavelength of the symmetron in vacuum, $\ell_{\infty}$. The grey regions mask values of $\mu$ and $M$ for which the best fit value of $v$ results in a fifth force that is  weaker than gravity: $\alpha=(v/M_{\rm Pl})^2(M_{\rm Pl}/M)^4<1$, or is incompatible with the predictivity of the Effective Field Theory (EFT) description: $v>M$. We also show as a dashed black line the boundary above which $v$ will always lead to a measurable fifth force in the Solar System, with the region below this line being permitted by the Cassini measurement. The white triangle indicates a particularly interesting point in the parameter space, where the symmetron fifth force explaining the motion of stars could be on the verge of detection with Solar System tests of gravity. The degeneracy in the parameter space where the effective mass of the symmetron field tends to zero can be seen running towards the lower right-hand corner of the plot. Figure reproduced with permission from Ref.~\cite{OHare:2018ayv}.}
	\label{fig:ciaran}
\end{figure}

As the scalar force is sourced by the baryonic matter in the galaxy, which forms the galactic disk, the symmetron profile perpendicular to the galactic plane may not have the same form as the radial dependence in the disk. This is in contrast to the distribution of particle dark matter around the galactic disk, which is expected to be approximately spherical.  As a result, the motion of stars perpendicular to the plane of the Milky Way could be used to discriminate between models.  In Ref.~\cite{OHare:2018ayv}, it was shown that the symmetron can be compatible with the perpendicular motion of stars in our local neighbourhood and still be sufficiently screened in the Solar System to satisfy PPN bounds. The results of a fit to the mock data set of Ref.~\cite{Read:2014qva} are shown in Fig.~\ref{fig:ciaran}. The degeneracy in the parameter space clearly runs  to smaller values of $M$ and larger values of $\mu$.
  The accuracy of  measurements of the motion of local stars will improve with new data coming from the Gaia satellite in the near future, and it may then be possible  to discriminate between the fifth-force and dark matter hypotheses directly. 

\section{Can the symmetron explain the lens mass of galaxies?}
\label{sec:lens}
If the symmetron model is to explain the rotation curves of disk galaxies entirely via an additional fifth force, its contribution to the total energy density of the galaxy must be subleading. This ensures that the scalar field configuration sources only a negligible increase in the depth of the gravitational potential of the galaxy over that due to the baryons.

In general relativity, gravitational lensing occurs because photon geodesics are curved by a  potential well.  The depth of the potential well is directly proportional to the mass of the galaxy, and so, in the thin-lens approximation, the deflection of a light ray is proportional to the `lens mass' - the mass of the galaxy. 
In observations of gravitational lensing, the baryonic mass of the galaxy is not sufficient to account for the total lens mass \cite{Massey:2010hh}. 
It is therefore clear that a symmetron fifth-force explanation of galactic rotation curves  will be incompatible with observations of gravitational lensing, unless there is some equivalent `fifth-force' effect acting on photons as they pass through the neighborhood of the galaxy.

\subsection{Perturbations to the metric potential}
\label{sec:potential}
We begin by checking  the consistency of our preceding assumption that, in the region of parameter space where the symmetron fifth-force can explain galactic rotation curves,  the energy density stored in the symmetron field has a negligible impact on the  curvature of spacetime.  As  a corollary, we demonstrate  that the store of energy in the scalar field in this regime cannot be the explanation for the difference between the lens mass and baryonic mass of a galaxy. 

The energy density stored in the scalar field is 
\begin{equation}
E_{\varphi}=\int {\rm d}V\; \left[\frac{1}{2}\left(\frac{\partial \varphi}{\partial r}\right)^2+\frac{1}{2}\left(\frac{\partial \varphi}{\partial z}\right)^2+\frac{\varphi^2\rho(r,z)}{2M^2}+U(\varphi)\right]\;,
\end{equation}
where
\begin{equation}
U(\varphi) = -\frac{1}{2}\mu^2\varphi^2 +\frac{\lambda}{4}\varphi^4+\frac{1}{4} \mu^2v^2\;,
\end{equation}
and we have shifted the zero of the potential to ensure that the symmetron contributes zero energy density (classically) to the cosmological vacuum, i.e.~$U(v)=0$.  

Assuming that the galaxy is cylindrically symmetric and using the equation of motion for $\varphi$, we find that
\begin{eqnarray}
E_{\varphi}&=&2\pi\int {\rm d}z\;\int r \; {\rm d}r\;\left[U(\varphi)-\frac{\varphi}{2}U^{\prime}(\varphi)\right]\\
&=& 2 \pi h r_s^2 \frac{\mu^2v^2}{4}\int_0^{\infty}{\rm d}x\; x \left(1-\left(\frac{\varphi}{v}\right)^4\right)\;.
\end{eqnarray}
In the last line, we  have redefined $x=r/r_s$ and  assumed that the the disk of the galaxy has a uniform height $h$. We have neglected  the variation in the $\varphi$ field perpendicular to the plane of the disk, instead assuming  that the field  is approximately constant over the height of the disk and zero outside it. This may  underestimate the amount of energy stored in the field, however we believe it captures the leading-order behaviour. 

The ratio of the energy stored in the scalar field to the mass of the galaxy is therefore
\begin{equation}
\frac{E_{\varphi}}{M_{\rm gal}}=\frac{1}{4} \frac{\mu^2M^2}{\bar{\rho}}\frac{v^2}{M^2}I\;,
\end{equation}
where $I=\int_0^{\infty}{\rm d}x\;x(1-(\varphi/v)^4)$ is a dimensionless constant, which depends on how `quickly' the field recovers to its vev, and $\bar{\rho}$ is the average three-dimensional density of the galaxy. For field profiles  such as those in Eq.~(\ref{eq:phiint}), which  approximate the form needed to explain galaxy rotation curves, we find that the integral $I$ evaluates to an order one number. If we require $v/M<1$ to be sure that the effective field theory is under control and if the galaxy is at least partially screened, so that $\bar{\rho}/\mu^2M^2 \geq 1$ then the amount of energy stored in the scalar field configuration can be neglected compared to the mass of the galaxy. 

This does, however, suggest that if we were working with a  theory where configurations with  $v \sim M$ remain well defined, and  if the critical density of the theory, $\rho_{\rm crit} = \mu^2M^2$ is the average galactic density then the energy stored in the scalar field would be comparable to the mass of the galaxy.   In this case, the scalar field profile would contribute directly to  the `missing mass' in the galaxy.  This is much closer to a typical scalar dark matter configuration, and the system would need to be reanalyzed to account for both the effects of the fifth force mediated by the scalar field and the contribution of the scalar to the total mass of the galaxy. Intriguingly, this choice of $\mu$ and $M$ is that picked out by the degenerate region in Fig.~\ref{fig:ciaran}. We will return to comment further on this intriguing possibility  later in this article.

\subsection{Axion-like couplings}
\label{sec:axionlike}
Classically, the photon kinetic term is invariant under conformal transformations\footnote{The square root of the metric determinant rescales as $\sqrt{-g} \rightarrow A^4(\varphi) \sqrt{-g}$, and the inverse metric as $g^{\mu\nu} \rightarrow A^{-2}g^{\mu\nu}$ so that $\sqrt{-g}g^{\mu\nu}g^{\rho\sigma}F_{\mu\rho}F_{\nu\sigma}\rightarrow\sqrt{-g}g^{\mu\nu}g^{\rho\sigma}F_{\mu\rho}F_{\nu\sigma}$.} and so the rescaling of the metric by the symmetron field in Eq.~(\ref{eq:scalarcoupling}) does not introduce a coupling between the symmetron scalar and photons.  However, viewing this model as a quantum field theory, it can be shown that quantum corrections generate an `axion-like' coupling between the scalar and photons \cite{Brax:2010uq}
\begin{equation}
\mathcal{L}\supset \frac{\varphi^2}{M_{\gamma}^2}\tilde{g}^{\mu\rho}\tilde{g}^{\nu \sigma}F_{\mu\nu}F_{\rho\sigma}\;,
\label{eq:alp}
\end{equation}
where $M_{\gamma}$ is a new energy scale, which is not fixed to be the same as the strength of the coupling to matter.  
This term  arises  because the conformal invariance of the photon Lagrangian is broken by a quantum anomaly that comes from integrating out  high-energy modes of the fermions  \cite{Brax:2010uq,Kaplunovsky:1994fg}.

\begin{figure}[t]
\centering
        \includegraphics[width=0.5\textwidth]{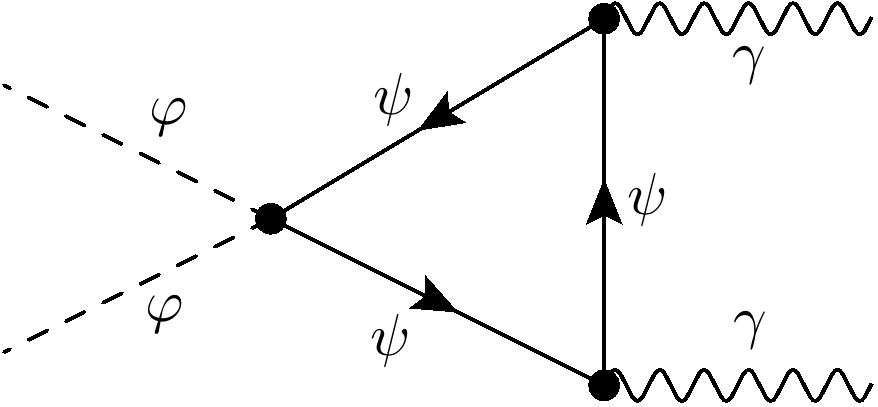}
  \caption{Feynman diagram showing how an axion-like coupling between photons and a scalar field can be mediated by a triangle loop of a fermionic field.}
	\label{fig:tri}
\end{figure}

Additionally, if there are heavy charged fermions beyond the Standard Model that couple directly to the scalar, they will be able to mediate interactions between the scalar and photons through a triangle loop diagram as in Fig.~\ref{fig:tri}. Integrating these heavy fermions out, leaves a low-energy effective theory that possesses a contact interaction between the conformally coupled scalar and two photons of the form in Eq.~(\ref{eq:alp})  \cite{Brax:2010uq,Nitti:2012ev}.

The coupling in Eq.~(\ref{eq:alp}) is known as an axion-like coupling, due to the similarities to the couplings of the hypothetical axion that solves the strong CP problem of QCD. Notice, however, that it is parity-even in this case. Including  this interaction means that the symmetron scalar field can modify the propagation of photons. In vacuum, the symmetron can now give rise to loop corrections to the photon propagator.  Additionally, in the presence of a background magnetic field, it can lead to non-conservation of photon number, and a change in the speed and polarisation of propagating photons.  We  consider each of these effects in turn.

\begin{figure}[t]
\centering
        \includegraphics[width=0.5\textwidth]{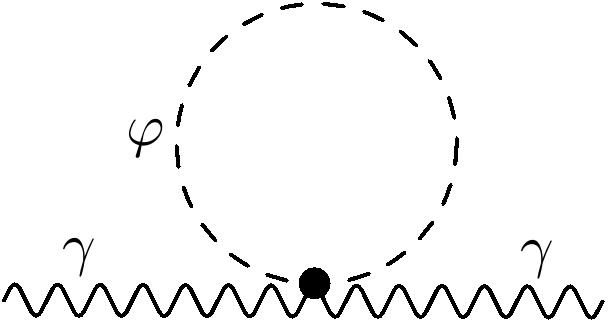}
  \caption{Feynman diagram showing the one-loop correction to the photon propagator due to the axion-like coupling of Eq.~(\ref{eq:alp}). In the symmetry broken phase for the symmetron model, there is an additional one-loop diagram which enters proportional to $(v/M)^2$, so it is suppressed by a higher power of $M$. }
	\label{fig:pol}
\end{figure}

The trace of the vacuum polarisation induced in the photon propagator by the loop of symmetron scalar in Fig.~\ref{fig:pol} is 
\begin{equation}
\Pi(k) = \frac{m_{\varphi}^2 k^2}{16 \pi M_{\gamma}^2}\ln\left(  \frac{\Lambda^2}{m_{\varphi}^2}\right)\;,
\end{equation}
where $\Lambda$ is an ultra-violet cut-off scale \cite{Brax:2009ey}, and $m_{\varphi}^2=V^{\prime\prime}_{\rm eff}(\varphi)$ is the local mass of the symmetron field. This is identically zero on-shell ($k^2=0$), as required by gauge invariance, although it will  have an effect on off-shell photons. As a result, the vacuum polarisation cannot give rise to the order-one correction to the  deflection angle needed to account for the observed lens mass of a galaxy.  

In a background magnetic field, the coupling in Eq.~(\ref{eq:alp}) introduces an effective operator that allows a photon to convert into two scalars  and vice versa.  In the presence of both a background magnetic field and a non trivial background scalar configuration, as we expect to find in a galaxy,  the effective operator describes mixing between one photon and a single fluctuation of the scalar field:
\begin{equation}
\mathcal{L}_{\rm eff}\supset \frac{4 B v }{M_{\gamma}}\frac{\varphi}{M_{\gamma}}(\partial_y A_x-\partial_x A_y)\;,
\label{eq:alp2}
\end{equation}
where we have chosen our coordinates so that the magnetic field $B$ is oriented along the $z$ direction, and $A_{\mu}$ describes the propagating photon.  The strength of the mixing between the photon and the symmetron scalar depends on both the strength of the background magnetic field and on $v/M_{\gamma}$.  This is the well-known  Primakov effect, which is also used to search for `invisible' axions \cite{Sikivie:1983ip}.  Depending on the orientation of the magnetic field, the propagating eigenstates of the system are a mix of one polarisation of the photon and the scalar field, and photon number is no longer conserved.   As the symmetron scalar field has a mass, the  mode which is a mix of the scalar and the photon will travel on a time-like geodesic.

In a constant and uniform magnetic field, only one polarisation of the photon mixes with the scalar, that polarised perpendicular to the orientation of the magnetic field.   The second  polarisation follows null geodesics.  In an unrealistic scenario  where a galaxy possesses a uniform magnetic field, mixing between the photon and a scalar field would produce two lensed images, one for each polarisation. The component of the photon that couples to the scalar will be lensed as if it were a particle with mass
\begin{equation}
m_{\rm eff}^2
 = \frac{v B \omega}{2 M_{\gamma}^2 }\;.
\end{equation}
Considering optical photons $\omega \sim \mbox{ eV}$, galactic magnetic fields of $B\sim 10\; \mu \mbox{G}$, $v/M_{\gamma} \sim 10^{-2}$ and a scalar photon coupling of $M_{\gamma} \sim M_{\rm Pl}/10$, this gives an effective mass of $m_{\rm eff} \sim 10^{-27}\mbox{ GeV}$.  To give rise to an order-one change in the deflection angle, the effective mass of the photon would have to become so large that it would become non-relativistic. This is not possible for light axion-like particles given the  current bounds on $M_{\gamma}$. Even allowing for the effects of the field to be screened in regions of high density, we require $M_{\gamma} \gtrsim 10^{9} \mbox{ GeV}$ \cite{Pettinari:2010ay,Burrage:2009mj,Burrage:2008ii}.

The magnetic fields of real galaxies are much more complex than this:  they may be split into coherent domains or display a turbulent structure.  Radio imaging of Faraday rotation also shows that the magnetic field extends beyond the location of the visible baryonic matter \cite{Brandenburg:2004jv}. This structure does mean that, in more realistic situations, both components of the photon will be mixed with the scalar field.  The biggest challenge for such an explanation of the observed lensing, however, is the need for the magnetic field strength to be directly correlated with the total mass of the galaxy, regardless of any other properties.  It is  already known that this is not the case; radio-faint galaxies have lower magnetic fields than gas rich galaxies with higher star formation rates \cite{Beck:2013bxa}. 

We are forced to conclude that the presence of an axion-like coupling for the symmetron scalar does not give rise to sufficient modifications to the photon geodesics to explain observations of gravitational lensing in the absence of  dark matter.

\subsection{Photon Mass}
In pure general relativity, the lensing deflection angle for a massive, rather than a massless, photon is \cite{PhysRevD.8.2349}
\begin{equation}
\Theta_{\rm total}= \frac{4 G \mathcal{M}_0}{b}\left( 1 + \frac{m_{\gamma}^2}{2 \nu^2}\right)\;,
\end{equation}
where $\mathcal{M}_0$ is the mass of the lens and $b$ is the impact parameter. A constant mass for the photon in general relativity is  tightly constrained to be $m_{\gamma} < 1 \times 10^{-18} \mbox{ eV}$ from considering the impact of the solar wind on the orbit of planets in the Solar System \cite{0741-3335-49-12B-S40,PhysRevD.98.030001}.  

Going beyond general relativity, we see that giving the photon a mass  means that the photon will now have a mass-dependent coupling to the symmetron scalar field through the universal coupling to the trace of the energy-momentum tensor (to which the photon mass will now contribute). The resulting equation describing the motion of the photon on a lensed trajectory is, to first order in the Schwarzschild radius $r_*=2G\mathcal{M}_0$,  
\begin{equation}
\frac{d^2 u}{d \theta^2} + u = \frac{r_*}{2}\left( 3 u^2 +  \frac{1}{J^2}-\frac{2AA^{\prime}\varphi_{,u}}{r_*J^2}\right)\;,
\label{eq:conformalmass}
\end{equation}
where $u=1/r$, $\theta \in (-\pi/2, \pi/2)$, $A^{\prime}=dA/d\varphi$ and $J$ is the conserved angular momentum. 
We have assumed that the correction to the photon trajectory due to the symmetron field is the same order in the expansion as the correction due to the Schwarzschild potential. This has to be true if the effects of the symmetron are to account for the observed deflection of light around galaxies. 

If the thin-lens approximation holds, we can assume that the photon trajectory is deflected only in a thin plane around the lens. It remains to estimate the form of the symmetron field around the source.  If the symmetron field is small compared to its vev at the point at which the photon trajectory crosses the lens plane then we can approximate $\varphi$ in Eq.~(\ref{eq:conformalmass}) as  
\begin{equation}
\varphi(u) = \lambda_0 v r_* u \left(\frac{M_{\rm Pl}}{M}\right)^2 \ll v\;,
\label{eq:smallerv}
\end{equation}
where  $\lambda_0$ is a screening factor which depends on the baryonic distribution in the lens.  In this case, 
\begin{equation}
A(\varphi)A^{\prime}(\varphi)\varphi_{,u} \approx  \lambda_0^2 r_*^2 u \left(\frac{v}{M}\right)^2 \left( \frac{M_{\rm Pl}}{M}\right)^4\;,
\end{equation}
and the correction due to the presence of the symmetron field will appear first  at order $r_*^2$. (The combination $\lambda_0 r_* u (M_{\rm Pl}/M)^2$ must remain small to ensure that the inequality in Eq.~(\ref{eq:smallerv}) remains satisfied.)

Alternatively, we can approximate the symmetron field as remaining close to its vev, such that
\begin{equation}
\varphi(u) = v -\lambda_0 v r_* u \left(\frac{M_{\rm Pl}}{M}\right)^2 \approx v\;,
\label{eq:nearlyv}
\end{equation}
where, again, $\lambda_0$ is a screening factor which depends on the lens. We note that the photon trajectory crosses the lens plane well outside the Schwarzschild radius of the galaxy, and so $r_*u\ll 1$. In this case, we find that 
\begin{equation}
A(\varphi)A^{\prime}(\varphi)\varphi_{,u} \approx - \lambda_0 r_*  \left(\frac{v}{M}\right)^2 \left( \frac{M_{\rm Pl}}{M}\right)^2\;,
\end{equation}
and the corrections due to the symmetron  enter the equation for the photon geodesic at order $r_*$, as required. The resulting deflection angle is 
\begin{equation}
\Theta_{\rm total} = \frac{4 G\mathcal{M}_0}{b}\left[ 1 + \frac{1}{2}\left(\frac{m_{\gamma}}{\nu }\right)^2 \left(1 +2\lambda_0\frac{  v^2M_{\rm Pl}^2}{ M^4}\right)\right]\;.
\end{equation}
The constraints on the mass of the photon, described above, mean that the additional correction to the deflection angle coming from the coupling to the symmetron is always subdominant to the general relativistic  result, and certainly cannot be responsible for the large inferred lens mass of  galaxies. (Again, the combination $\lambda_0 (M_{\rm Pl}/M)^2$ must remain small to ensure the validity of the approximation in  Eq.~(\ref{eq:nearlyv}).)

One final possibility remains: that the photon gets an `effective mass' from a coupling to the symmetron scalar, which varies depending on the environment. For example, the symmetron could give the photon a mass through a Higgs-like mechanism, i.e.~
\begin{equation}
\mathcal{L} \supset \tilde{\alpha}\varphi^2A_{\mu}A^{\mu}\;,
\label{eq:photoncoup}
\end{equation}
where $\tilde{\alpha}$ is a dimensionless coupling constant.  This  operator  would mean that the photon mass would vanish locally, and in regions of high density where $\varphi \sim 0$, and would only be relevant on galactic scales where the field is weakly perturbed from its vev. This is the only regime in which we need to perturb the behavior of photons in order to  explain the lens mass of galaxies.

The constraints on the photon mass reported by the Particle Data Group \cite{Patrignani:2016xqp} come from local tests within the Solar System, but constraints also come from observations of galactic magnetic fields \cite{Chibisov:1976mm,Adelberger:2003qx} on much larger distance scales.  A mass for the photon means that there is an additional contribution to the energy budget of the galaxy, which is proportional to the photon mass.  Taking into account the effect that this energy has on the galactic plasma leads to the bound $m_{\gamma}\lesssim 10^{-27} \mbox{ eV}$ \cite{Lakes:1998mi}, and studies of the large-scale magnetic field of the galaxy and the gas pressure of the galaxy imply $m_{\gamma} \lesssim 10^{-26} \mbox{ eV}$ \cite{Chibisov:1976mm}.  These bounds depend on the origin of the photon mass, however, and it was shown in Ref.~\cite{Adelberger:2003qx}  that if the photon mass arises through a Higgs mechanism then the bounds from galactic physics can be avoided altogether. It is for  this reason that the bounds from galactic physics are not used by the Particle Data Group.    The constraints are avoided in such a model if the energy density in the galactic magnetic field is sufficiently large that it restores the symmetry for the scalar field, driving the scalar vev and, consequently, the photon mass to zero. There is an additional regime in which the symmetry is not restored across the whole galaxy, but a network of vortices balances out the energy due to the massive photon field.  (We refer the reader to Ref.~\cite{Adelberger:2003qx} for further details of this mechanism.) 
A symmetron model that gives rise to a photon mass through a Higgs mechanism  cannot avoid the bounds on the photon mass from galactic physics.  If the symmetry is restored in a sufficiently large region of the galaxy to avoid the constraints on the photon mass from galactic magnetic fields,  the field profile needed for the symmetron to explain the rotation curves of galaxies will be destroyed. As the effective mass of the symmetron is of order the size of the galaxy, it is also not possible to create a sufficiently dense network of symmetron topological defects to cancel the photon mass contribution to the galactic energy budget.

Constraints on the photon mass also arise from cosmological observations of gamma ray bursts \cite{Schaefer:1998zg}, extra-galactic pulsars \cite{Wei:2018pyh,Wei:2016jgc} and fast radio bursts \cite{Bonetti:2016cpo,Wu:2016brq,Bonetti:2017pym}, which give the tightest constraint $m_{\gamma} \lesssim 10^{-14} \mbox{ eV}$. These constraints are significantly  weaker than those obtained from cosmological measurements.  We could, however, ensure that any bound on the photon mass was always avoided by generating the mass through an operator of the form 
\begin{equation}
\mathcal{L} \supset \frac{\varphi^2}{M_{\gamma}^{\prime\;2}}(v^2-\varphi^2)A_{\mu}A^{\mu}\;,
\label{eq:photoncoup2}
\end{equation}
where $M_{\gamma}^{\prime}$ is the constant mass scale controlling this interaction. Although this is clearly a tuned choice, the form of the interaction in Eq.~(\ref{eq:photoncoup2}) would ensure that there was no coupling in the cosmological vacuum where $\varphi \sim v$.  In the remainder of this section, we consider only the simpler form for the symmetron photon coupling of Eq.~(\ref{eq:photoncoup}), but the two possibilities will behave very similarly at linear order in the perturbed $\varphi$.

 To find the corrections to the geodesic equation introduced by a Higgsed mass for the photon of the form in Eq.~(\ref{eq:photoncoup}), we vary the action for a single massive particle 
\begin{equation}
S= - \int \alpha(\varphi) \;{\rm d}s\;,
\end{equation}
where $\alpha(\varphi)= \tilde{\alpha}^{1/2}\varphi$ is the effective photon mass,\footnote{This $\alpha$ is not to be confused with the $\alpha$ appearing earlier in Section \ref{sec:intra} that parametrizes the strength of the fifth force for matter. } where $s$ is the proper time along the particle's geodesic.
To first order in the Schwarzschild radius, we find that the resulting geodesic equation is 
\begin{equation}
\frac{d^2 u}{d \theta^2} + u = \frac{r_*}{2}\left( 3 u^2 +  \frac{\alpha^2(\varphi)}{L^2}\right)-\frac{\alpha^{\prime}(\varphi)\alpha(\varphi)\varphi_{,u}}{L^2}(1-ur_*)\;,
\end{equation}
where $L= (\alpha(\varphi)/u^2)(d\theta/ds)$ is a conserved quantity of the motion. (If the photon mass were fixed then $L/m_{\gamma}=J$ would be the conserved angular momentum.) 
The galactic bounds on the photon mass  imply that, at most, $\alpha(1/b) \sim 1/r_s$.  This means that the symmetron-induced correction to the geodesic equation can only be subdominant to the gravitational effects.   Assuming that the dominant corrections to the photon geodesic occur in the lens plane, and that $\alpha(1/b) \sim 1/r_s$, we find the lensing deflection angle is
\begin{equation}
\Theta_{\rm total} = \frac{4 G\mathcal{M}_0}{b}\left[ 1 + \frac{2}{(\pi \nu r_s)^2} \left(1 +2\lambda_0\frac{  M_{\rm Pl}^2}{ M^2}\right)\right]\;,
\end{equation}
where   we have again assumed the form for $\varphi$ given in Eq.~(\ref{eq:nearlyv}). We see that the correction due to the effects of the symmetron field can only be subdominant to the leading general relativistic term.

\subsection{Disformal coupling}

We  now make the simplest extension of  the symmetron model introduced in Refs.~\cite{Burrage:2016yjm,Burrage:2016xzz} by adding a disformal coupling term \cite{Bekenstein:1992pj}. This means that matter fields move on geodesics of the metric
\begin{equation}
g_{\mu\nu}= A^2(\varphi)\tilde{g}_{\mu\nu}+B(\varphi)\partial_{\mu}\varphi \partial_{\nu}\varphi\;,
\end{equation}
where we note that the function $B$", which should not be confused with the magnetic field strength appearing  in
Sec.~\ref{sec:axionlike}, has mass dimension $1/M_{\rm dis}^4$. In what follows, we  will assume that $B$ is independent of $\varphi$ at leading order.

It has been shown \cite{Bruneton:2007si} that, at first post-Newtonian order and assuming that the scalar field profile is static and spherically symmetric, the disformal coupling contributes to the lensing deflection angle as 
\begin{equation}
 \Theta_{\rm total} =  \Theta_{\rm GR} +\int_{b}^{\infty}\frac{b^2 \;{\rm d}r}{r\sqrt{r^2-b^2}}\frac{B(\partial_r\varphi)^2}{A^4}\;,
\end{equation}
where $b$ is the impact factor and $\Delta\theta_{\rm GR}= 4 G \mathcal{M}_0/b$ for a lens of mass $\mathcal{M}_0$.  Assuming that there is no dark matter in the galaxy, $\mathcal{M}_0$ is the total baryonic mass. We have set $c=1$ in this expression but, reintroducing $c$, we would see that corrections to this expression appear first at order $1/c^4$. 

If galactic rotation curves are explained by a symmetron fifth force then the form of the scalar field profile is given by Eq.~(\ref{eq:phiint}).
As a result, the deflection angle induced by a disformally coupled symmetron, relative to the deflection angle around the same mass in pure general relativity,  is 
\begin{equation}
\frac{ \Theta_{\rm symm}}{\Theta_{\rm GR}}= \frac{bM^4 g_{\dagger}^2}{4 G \mathcal{M}_0}\int_b^{\infty}\frac{{\rm d}r}{r}\frac{1}{\sqrt{r^2/b^2 -1}}\frac{B}{\varphi^2A^4}\frac{f^4(x)}{(e^{f(x)}-1)^2}\;.
\end{equation}
For consistency, we always work in the regime where $\varphi/M\ll 1$ and so we approximate $A \approx 1$ and set $\varphi^2 \sim v^2$ to find
\begin{equation}
\frac{\Theta_{\rm symm}}{ \Theta_{\rm GR}}= \left(\frac{M}{v}\right)^2(B M^4) \frac{1}{M^2}\frac{b}{r_s}\frac{g_{\dagger}}{r_s }\frac{1}{4 f_0}\int_{b/r_s}^{\infty}
\frac{{\rm d}x}{x}\frac{1}{\sqrt{x^2 r_s^2/b^2-1}}\frac{f^4(x)}{(e^{f(x)}-1)^2}\;.
\end{equation}

We take the characteristic values $f_0 =5$ and $b/r_s = 100$  to find
\begin{equation}
\frac{\Theta_{\rm symm}}{\Theta_{\rm GR}}\approx 2 \times 10^{-3} \left(\frac{M}{v}\right)^2 (B M^4) \frac{g_{\dagger}}{r_s  M^2}\;.
\end{equation}
This means that the contribution to the deflection angle from the symmetron field is of the same order as the general relativistic contribution when 
\begin{equation}
B \approx \left(\frac{10^{29}}{M_{\rm Pl}}\right)^4\;,
\end{equation}
where we have used $g_{\dagger} = 1.2 \times 10^{-10} \mbox{ ms}^{-2}$, c.f.\ Eq.~(\ref{eq:relrel}),and the parameters $r_s=5 \mbox{ kpc}$,  $v/M=1/150$ and $M=M_{\rm Pl}/10$ for which the symmetron can explain the rotation curves of disk galaxies, as discussed in Section \ref{sec:intra}.
This requires the energy scale of the disformal coupling to be  $M_{\rm dis} \sim 10 \mbox{ meV}$,  a value excluded by terrestrial experiments \cite{Brax:2014vva,Brax:2015hma}, which require the energy scale of the disformal coupling to be greater than $650 \mbox{ GeV}$.

\section{Conclusions}
\label{sec:conc}

Light scalar fields coupled to gravity have been proposed as both theories of dark matter and theories of modified gravity.  In this work, we consider such a scalar to behave as dark matter on galactic scales if it gives rise to a significant contribution to the total mass of the galaxy. A scalar acts as modified gravity on galactic scales if it does not contribute significantly to the total mass of the galaxy but instead mediates a fifth force that modifies the motion of standard model particles. These are two extremes of the phenomenology that can be exhibited by a  single scalar field theory, and so a mixed regime where the scalar contributes to the total mass and mediates a fifth force should also be considered as a possibility within these simple theories. 

The symmetron is an example of such a theory that introduces a single real scalar field, that couples non-minimally to matter (or to gravity depending on the choice of frame).  The scalar has a Higgs-like potential and couples to matter in such a way that the $\varphi \rightarrow -\varphi$ symmetry is restored in regions of high density and broken in regions of low density.  Due to this coupling to matter, the symmetron mediates a fifth force, the strength of which depends on the local value of the scalar field. 

It has previously been shown that a symmetron-mediated fifth force could explain the internal dynamics of galaxies: their rotation curves, stability (to the formation of bars), and the motion of stars perpendicular to the galactic plane. In this work,  we asked whether such a theory with a single symmetron field  can explain the difference between the baryonic mass and the lens mass of a galaxy purely as a modification of gravity. 
 We have considered the conformal coupling of the scalar, the contribution of the energy stored in the scalar profile to the total mass of the galaxy, the axion-like couplings induced by quantum effects, a possible mass of the photon and inclusion of an additional disformal coupling. To the best of our knowledge, this comprises a complete survey of the effects of a single universally coupled real scalar field. We conclude that it is not possible for a single  symmetron field to explain the lensing of photons around galaxies purely through a modification of gravity.

Without further extending the field content beyond the symmetron model, we see one possibility for the symmetron to continue to explain galactic dynamics and the lensing of light by galaxies: that  a combination of modified gravity and dark matter effects are relevant on galactic scales.   As shown in Section \ref{sec:potential}, this can occur when the critical density for the symmetron is of order the average galactic density and when $v \sim M$.  This  requires a reanalysis of the data, where both the symmetron fifth force, and its correction to the galactic Newtonian potential are taken into account.  The combined regime offers an intriguing possibility to capture the successes of both the particle dark matter and modified gravity paradigms within one simple scalar model.  However,  it pushes the  theory into the non-linear regime,  making the observational consequences of the symmetron more difficult to compute.  We intend to investigate the dark matter - modified gravity symmetron further in future work.

Alternatively, one can further extend the field content. An example of this approach is given in  Ref.~\cite{Khoury:2014tka}, where a related attempt was made to explain galactic rotation curves with a scalar fifth force, in this case with a modified kinetic term. Therein, a second scalar field $\pi$ could be introduced,
 which couples to matter disformally as 
\begin{eqnarray}
g_{00}&=& \tilde{g}_{00} + \frac{\varphi^2}{W^2}\partial_0 \pi \partial_0 \pi,
\end{eqnarray}
where $\varphi$ is given in Ref.~\cite{Burrage:2016xzz}, $W$ is a constant with dimensions of mass  and $\pi$ is a scalar field, whose time evolution is specified so that $\pi' \approx 1$, where $'$ indicates a derivative with respect to
conformal time. It has been suggested that this second scalar field could be responsible for the contribution of dark matter to the background evolution of the universe, and its disformal coupling for the additional lens mass of galaxies.

\section*{Acknowledgements}
We would like to thank Ciaran O'Hare, Daniela Saadeh, and Ben Elder for useful conversations during the course of this work. CB and PM are supported by a Leverhulme Trust Research Leadership Award.  CB is also supported in part by a Royal Society University Research Fellowship. EJC would like to acknowledge financial support from STFC Consolidated Grant No. ST/P000703/1. CK acknowledges support by the University of Nottingham and a University of Nottingham Vice Chancellor's Scholarship for Research Excellence.

\end{document}